\documentclass[%
reprint,
superscriptaddress,
showpacs,preprintnumbers,
 amsmath,amssymb,
 aps,
]{revtex4-1}

\usepackage{graphicx}
\usepackage{float}

\newcommand{\myfig}[4][ht]{
\begin{figure}[#1]
\centering
\includegraphics[#2]{#3}
\caption{#4\label{#3}}
\end{figure}
}

\graphicspath{{images/}}

\usepackage{dcolumn}
\usepackage{bm}

\begin{document}
\title{Compact models for multidimensional quasiballistic thermal transport}
\author{Bjorn Vermeersch}%
 \email{Email: bjorn.vermeersch@cea.fr}%
\affiliation{CEA, LITEN, 17 Rue des Martyrs, 38054 Grenoble, France}%
\vspace{2ex}
\date{\today}
\begin{abstract}
The Boltzmann transport equation (BTE) has proven indispensable in elucidating quasiballistic heat dynamics. Experimental observations of nondiffusive thermal transients, however, are interpreted almost exclusively through purely diffusive formalisms that merely extract `effective' Fourier conductivities. Here, we build upon stochastic transport theory to provide a characterisation framework that blends the rich physics contained within BTE solutions with the convenience of conventional analyses. The multidimensional phonon dynamics are described in terms of an isotropic Poissonian flight process with rigorous Fourier-Laplace single pulse response $P(\vec{\xi},s) = 1/[s + \psi(\| \vec{\xi} \|)]$. The spatial propagator $\psi(\|\vec{\xi}\|)$, unlike commonly reconstructed mean free path spectra $\kappa_{\Sigma}(\Lambda)$, serves as a genuine thermal blueprint of the medium that can be identified in compact form directly from raw measurement signals. Practical illustrations for transient thermal grating (TTG) and time domain thermoreflectance (TDTR) experiments on respectively GaAs and InGaAs are provided.
\end{abstract}

\maketitle


\section{Introduction}
Thermal transport in dielectric solids over length scales comparable to phonon mean free paths (MFPs) deviates from standard diffusive predictions \cite{lengthscale}. The theoretical understanding of such `quasiballistic' heat dynamics, as well as the technical capabilities to observe them experimentally, have expanded extensively over the past decade \cite{TTG-fulldetails,TTG-GaAs,TTG-membranes,siemens,hoogeboom,nanograting,minnich-spotsize,wilson-anisotropy,cahill,malen,malenRSI,PRBlevy2,maznev-gratingtheory,MFPspectroscopy,collins,wilson-twochannel,minnichBTE1D,minnichBTE3D,minnichVRMC,regner,kohnonlocal,PRBlevy1,minnich-TTG3D}. Some measurements induce nondiffusive behaviour directly by altering the physical size of the heat source. Key examples of scalable sources include laser interference patterns used by transient thermal grating (TTG) \cite{TTG-fulldetails,TTG-membranes,TTG-GaAs}, metal nanogratings employed by soft x-ray metrology \cite{siemens,hoogeboom} and two-tint time domain thermoreflectance (TDTR) \cite{nanograting}, and TDTR laser spot diameter \cite{minnich-spotsize,wilson-anisotropy}. The characteristic length scale of the thermal gradient can also be varied indirectly, as is done in time/frequency domain thermoreflectance (TDTR/FDTR) through the pump laser modulation frequency \cite{cahill,malen,malenRSI,PRBlevy2,wilson-anisotropy}.
\par
The essential physics that underpin the quasiballistic heat flow observations have been theoretically explained by a variety of works \cite{maznev-gratingtheory,MFPspectroscopy,collins,wilson-twochannel,minnichBTE1D,minnichBTE3D,minnichVRMC,regner,kohnonlocal,PRBlevy1,minnich-TTG3D} largely indebted to the Boltzmann transport equation (BTE). However, BTE solutions are formulated in terms of a wide spectrum of phonon modes and as such are too intricate and inflexible for direct processing of experimental data. Instead, nearly all measurements of inherently nondiffusive heat dynamics are analysed using conventional diffusive theory and interpreted in terms of `effective' Fourier thermal conductivities/resistivities \cite{TTG-GaAs,TTG-membranes,siemens,hoogeboom,nanograting,minnich-spotsize,wilson-anisotropy,cahill,malen,malenRSI}. Several of our prior works \cite{PRBrms,PRBlevy2} have pointed out substantial artifacts induced by this methodology, and other community members have joined in expressing the growing need for `beyond Fourier' characterisation frameworks \cite{hoogeboom,maassen1}.
\par
A recent study \cite{maassen1,maassen2} suggested convenient tackling of quasiballistic transport through hyperbolic diffusion equations with carefully formulated boundary conditions. However, closer investigation \cite{PRBcomment} shows that this method fails to capture the inherent onset of nondiffusive dynamics at length scales comparable to phonon MFPs in periodic heating regimes, rendering the proposed hyperbolic framework ill suited to experimental characterisation.
\par
We have previously introduced a `truncated L\'evy' approach for refined interpretation of TDTR experiments on semiconductor alloys \cite{PRBlevy2}. This method is capable to extract the L\'evy superdiffusion exponent $\alpha$ that regulates the alloy's quasiballistic heat dynamics \cite{PRBlevy1} directly from raw measurement data and thereby offers comprehensive insight not accessible through effective Fourier theory. However, the model bears a significant computational cost because its geometric extension of 1D truncated L\'evy motion to 3D heat flow geometries entails multiple numerical integral transforms.
\par
Here, we develop an improved framework for quasiballistic characterisation of alloy and non-alloy compounds through rigorous analytic treatment of multidimensional isotropic stochastic processes. Within the theory, outlined in Sec. \ref{sec:theory}, a spatial propagator function $\psi$ emerges as thermal blueprint that can be described in compact form (Sec. \ref{sec:parametric}). The model eliminates several inaccuracies and inconveniences of its predecessor while offering superior computational efficiency (Sec. \ref{sec:improvements}).  Section \ref{sec:applications} illustrates practical applications to TTG and TDTR analyses. A brief summary (Sec. \ref{sec:summary}) concludes the paper.
\section{Theory}\label{sec:theory}
\subsection{Essential background: Poissonian flights}
Our approach is firmly rooted in stochastic transport theory \cite{montrollweiss1,klafter} and based on Poissonian flight processes. Here we briefly review key essentials. A flight process describes random transient motion in $d$-dimensional space through a series of transition events. A `jump driver' $\phi_d(\vec{r},\vec{r}')$ dictates the probability to relocate from $\vec{r}$ to $\vec{r}'$ while a stochastically independent distribution $\varphi(\vartheta)$ governs the `wait time' $\vartheta$ between consecutive jumps. For homogeneous media, the jump driver reduces to $\phi_d(\vec{r} - \vec{r}')$ and the process can be fully characterised by its `single pulse response' $P(\vec{r},t)$. This function describes the chance of finding a random wanderer in location $\vec{r}$ at time $t$ after it was released from the origin at $t=0$ and can be obtained in Fourier-Laplace domain ($\vec{r} \leftrightarrow \vec{\xi}$, $t \leftrightarrow s$) through the Montroll-Weiss equation \cite{montrollweiss1,montrollweiss2}:
\begin{equation}
P(\vec{\xi},s) = \frac{1 - \varphi(s)}{s\, \left[ 1 - \varphi(s) \, \phi_d(\vec{\xi}) \right]} \label{Pxis}
\end{equation}
In a thermal context, $P \equiv C_v \, \Delta T$ denotes the deviational volumetric energy density, where $C_v$ is the medium's heat capacity and $\Delta T$ the temperature rise relative to ambient. For the particular case of `Poissonian' processes, being those with exponentially distributed wait time $\varphi(s) = (1 + s \vartheta_0)^{-1}$, the solution takes the form
\begin{equation}
P(\vec{\xi},s) = \frac{1}{s + \psi_d(\vec{\xi})} \,\, \leftrightarrow \,\, P(\vec{\xi},t) = \exp \left[- \psi_d(\vec{\xi}) \, t \right] \label{Poissonian}
\end{equation}
where $\psi_d(\xi) \equiv [ 1 - \phi_d(\vec{\xi}) ]/\vartheta_0$. We will call $\psi_d$
the \textit{propagator function} since, as (\ref{Poissonian}) conveys, this entity describes the spatiotemporal propagation of thermal energy.
\subsection{Simplification to isotropic transport}
Realistic phonon dispersions $\omega(\vec{k})$ and scattering rates $\tau^{-1}(\vec{k})$ usually display directional dependences. Even so, thermal transport in many semiconductors, especially those with cubic/zincblende crystal structures, can be considered isotropic within good approximation \cite{minnichBTE3D,APLthinfilms}. In the stochastic context, isotropic motion arises through jump drivers $\phi_d(\|\vec{r}-\vec{r'}\|)$ that depend only on the distance between transition sites but not their relative spatial orientation. We can consequently exploit that a $d$-dimensional radially symmetric function $f_d(\|\vec{r}\|)$ has a radially symmetric Fourier image $F_d(\|\vec{\xi}\|)$ obtainable through a univariate integral transform \cite{fouriertransforms}. In 2D this property is embodied by the Hankel transform $f(\rho) \leftrightarrow F(h)$ with Bessel kernel $J_0(h \rho)$; the 3D pair $f(r) \leftrightarrow F(\zeta)$ involves the spherical Bessel kernel $j_0(\zeta r) \equiv \sin(\zeta r) / (\zeta r)$. (Explicit formulae are listed in Appendix \ref{app:fouriertransforms} for convenience.) Owing to these mathematical symmetries, Poissonian propagators $\psi_d$ for multidimensional isotropic transport are always expressable as univariate functions. These moreover relate rigorously to their 1D counterparts through a straightforward variable exchange $\xi_x \leftrightarrow \|\vec{\xi}\|$ (explicit proof in Appendix \ref{app:mappingproof}).
\subsection{Propagator functions as thermal blueprint}
A variaty of studies \cite{hoogeboom,nanograting,malen,malenRSI,MFPspectroscopy,collins,regner,minnichBTE1D,minnich-TTG3D} have reconstructed the `MFP spectrum' (cumulative conductivity function) $\kappa_{\Sigma}(\Lambda)$ from effective conductivities $\kappa_{\text{eff}}(\chi)$ measured as a function of a controllable parameter $\chi$. Although the $\kappa_{\Sigma}$ curve reveals the spatial extent of distinct transport regimes \cite{PRBlevy1}, it holds insufficient information for computing the actual quasiballistic heat dynamics. Moreover, $\kappa_{\Sigma}$ reconstruction requires solving inverse problems $\kappa_{\text{eff}}(\chi) = \int_{0}^{\infty} S(\Lambda,\chi) \, (\partial \kappa_{\Sigma} / \partial \Lambda) \, \mathrm{d}\Lambda$, where $S$ is a theoretical `suppression function' that depends on the particular details and geometry of the experiment. Here, we instead will perform direct parametric identification of the propagator function $\psi_d(\|\vec{\xi}\|)$, which does constitute a genuine thermal blueprint of the medium per Eq. (\ref{Poissonian}), from raw measurement signals. This not only bypasses the potential ambiguities associated with effective conductivities but also eliminates the need to determine the suppression function $S$. A firm connection between the `macroscopic' propagator and underlying `microscopic' phonon properties is moreover still maintained in the form \cite{PRBlevy1}
\begin{equation}
\psi_d = \sum \frac{C_k \, \| \vec{\xi} \|^2 \Lambda_{\hspace{-0.1em}/\hspace{-0.2em}/k}^2}{\tau_k (1 + \| \vec{\xi} \|^2 \Lambda_{\hspace{-0.1em}/\hspace{-0.2em}/ k}^2)} \, \biggr / \sum \frac{C_k}{1 + \| \vec{\xi} \|^2 \Lambda_{\hspace{-0.1em}/\hspace{-0.2em}/ k}^2} \label{psi_phonons}
\end{equation}
Subscripts $_{\hspace{-0.1em}/\hspace{-0.2em}/}$ indicate projections on a cartesian axis (note that all crystal directions are considered equivalent since we operate under isotropic assumptions); the summations over discrete wavevector space can be easily reformulated as integrals over phonon frequency.
\subsection{Note regarding transition velocity}
The stochastic independence between jump length and wait time in isotropic flight processes induces hopping trajectories with unregulated transition velocities. This might raise conceptual concerns towards thermal modelling, since phonons propagate at well defined group velocities. In practice, however, most experimental observations operate within the so called weakly quasiballistic regime $t/\tau \gg 1 \leftrightarrow |s| \tau \ll 1$ in which the Green's function $G(\xi,t)$ of the 1D BTE is known to obey an exponential time decay with $\xi$-dependent rate \cite{minnichBTE1D}. From a stochastic viewpoint, $G$ thus conforms precisely to the characteristic signature (\ref{Poissonian}) of a Poissonian flight process, as we also pointed out in prior first-principles work \cite{PRBlevy1}. After analytic solid angle integrations, the Green's function of the 3D BTE for an isotropic crystal was additionally found to become formally identical to its 1D counterpart except for variable exchange $\xi_x \leftrightarrow \| \vec{\xi} \|$ \cite{minnichBTE3D}. As we saw earlier, isotropic flight processes exhibit precisely the same property through generic mathematical symmetry.
\section{Parametric forms for $\psi_d(\|\vec{\xi}\|)$}\label{sec:parametric}
The accuracy and convenience of our framework hinges on utilising physically suitable propagator functions in compact form. One archetypical process family of particular interest here is that of the alpha-stable (L\'evy) flights. These are closely associated with fractional-space anomalous diffusion \cite{klafter,levyflights} and defined by \cite{levytheorems}
\begin{equation}
\psi_d(\|\vec{\xi}\|) = D_{\alpha} \, \|\vec{\xi}\|^{\alpha} \qquad 1 \leq \alpha \leq 2 \label{levyprocess}
\end{equation}
Here $\alpha$ is the characteristic exponent and $D_{\alpha}$ (unit m$^{\alpha}$/s) is the `fractional diffusivity'. Brownian motion (diffusive transport) corresponds to $\alpha = 2$; one easily verifies that inverse transform of $\exp(-D\, \|\vec{\xi}\|^2 \, t)$ indeed produces the Gaussian kernels $P(\|\vec{r}\|,t) = (4\pi D t)^{-d/2} \exp(\|\vec{r}\|^2/4Dt)$ of the Fourier heat equation. The case $\alpha = 1$, where $D_{\alpha}$ takes on the meaning of characteristic heat propagation velocity $\bar{v}$, produces Cauchy distributions $P(\|\vec{r}\|,t) = \bar{v} t/A_d\pi [\bar{v}^2\,t^2 + \|\vec{r}\|^2]^{(d+1)/2}$ where $A_d = \{1,2,\pi\}$ for $d=\{1,2,3\}$. Solutions for intermediate L\'evy exponents are expressable as infinite power series in $\|\vec{r}\|/(D_{\alpha}\,t)^{1/\alpha}$ \cite{klafter} and possess the following generic properties \cite{levytheorems}:
\begin{eqnarray}
P(\|\vec{r}\| = 0,t) & \sim & (D_{\alpha} \, t)^{-d/\alpha} \quad 1 \leq \alpha \leq 2 \label{levysource}\\
P(\|\vec{r}\| \gg [D_{\alpha} \, t]^{1/\alpha},t) & \sim & \|\vec{r}\|^{-(d+\alpha)} \,\,\quad 1 \leq \alpha < 2 \label{levytails}
\end{eqnarray}
\subsection{Alloy compounds}
We have previously shown \cite{PRBlevy1} that phonon scattering mechanisms of the form $\tau \sim \omega^{-n} (n>3)$, such as ideal mass impurity scattering $n=4$, naturally induce L\'evy dynamics with $\alpha = 1+3/n$. L\'evy flights, therefore, are highly relevant to model quasiballistic thermal transport in semiconductor alloys. Pure L\'evy processes, however, maintain their hallmark characteristics (\ref{levysource}) and (\ref{levytails}) indefinitely because the jump driver $\phi_d$ has infinite variance. Phonon MFPs, by contrast, are always physically bounded through either boundary scattering or macroscopic dissipation effects \cite{longMFPs}. This restores a finite jump length variance which inherently ensures recovery to Brownian motion at long length and time scales \cite{montrollweiss1}. We can describe the complete behaviour by
\begin{equation}
\psi_d(\|\vec{\xi}\|) = \frac{D \, \|\vec{\xi}\|^2}{(1 + r_{\text{LF}}^2 \, \|\vec{\xi}\|^2)^{1-\alpha/2}} \label{temperedlevypropagator}
\end{equation}
This `tempered L\'evy' process evolves from alpha-stable dynamics with fractional diffusivity $D_{\alpha} = D/r_{\text{LF}}^{2-\alpha}$ to Fourier diffusion over characteristic length scale $r_{\text{LF}}$. We note that nearly identical transitions can also be described by relativistic stable processes $\psi_d(\|\vec{\xi}\|) = D_{\alpha} [(\| \vec{\xi} \|^2 + M^{2/\alpha})^{\alpha/2} - M]$ \cite{relativisticlevy} with `mass' $M = (\alpha/2)^{\alpha/(2-\alpha)}/r_{\text{LF}}^{\alpha}$ (unit 1/m$^{\alpha}$). However, our custom formulation (\ref{temperedlevypropagator}) conveniently describes the Levy-Fourier transition through an easy-to-interpret lengthscale parameter $r_{\text{LF}}$ closely related to characteristic MFPs. Mathematically, the instantaneous propagator exponent $\partial \ln \psi_d / \partial \ln \| \vec{\xi} \|$ passes through its midpoint $(2+\alpha)/2$ precisely at $\| \vec{\xi}\| = r_{\text{LF}}^{-1}$ and achieves 90\% of its total swing over the interval $1/3 \leq r_{\text{LF}} \, \|\vec{\xi}\| \leq 3$.
\subsection{Non-alloy compounds (`single crystals')}
Single crystals typically do not display L\'evy dynamics because of their smaller scattering exponents $n \simeq 3$. At this pivotal value, power-law dependences for MFP spectra and cross-plane film conductivities $\kappa_{\Sigma,\perp} \sim \{\Lambda,L\}^{2-\alpha}$ observed for alloys turn into logarithmic ones: $\kappa_{\Sigma,\perp} \sim \ln \{\Lambda,L\}$ \cite{PRBlevy1,APLthinfilms}. First-principles calculations indicate that the reduced propagator $\tilde{\psi}_d \equiv \psi_d/D\|\vec{\xi}\|^2$ in single crystals exhibits a similar logarithmic transition between its diffusive ($\tilde{\psi}_d \simeq 1$) and quasiballistic ($\tilde{\psi}_d \sim \|\vec{\xi}\|^{-1}$) asymptotes. We can therefore propose the following `log-tempered' parametric form:
\begin{equation}
\psi(\|\vec{\xi}\|) = \frac{D \, \|\vec{\xi}\|^2}{\ln(2)} \, \ln \left[ 1 + \left(1 + r_{\text{CF}}^{b} \, \|\vec{\xi}\|^{b} \right)^{-1/b} \right] \label{logtemperedcauchypropagator}
\end{equation}
This process evolves from Cauchy dynamics with velocity $\bar{v} = D/[\ln(2) \, r_{\text{CF}}]$ to regular Fourier diffusion over characteristic length scale $r_{\text{CF}}$, with large/small exponents $b \gtrless 1$ signifying a sharp/broad quasiballistic transition.
\section{Improvements over prior work}\label{sec:improvements}
In earlier work we introduced a `truncated L\'evy' model for quasiballistic TDTR analysis of semiconductor alloys \cite{PRBlevy2}. This theory was based on a 1D Poissonian flight process that was geometrically extended to 3D heat flow in real space/time domain: 
\begin{equation}
\text{prior work: } P_{\text{geo}}(r,t) = P^3_{\text{1D}}(x = r/\sqrt{3},t) \label{geoextension}
\end{equation}
This approach suffers from several drawbacks. First, hindsight revealed that the geometric extension (\ref{geoextension}) is not exact for non-Brownian motion. Pure L\'evy transport, for example, produces improperly normalised distributions $\iiint P_{\text{geo}} \mathrm{d}\vec{r} < 1$ with correct source transient signature $P_{\text{geo}}(0,t) \sim t^{-3/\alpha}$ but incorrect spatial decay (tail exponent $3+3\alpha$ instead of $3+\alpha$). These characteristics allowed proper $\alpha$ identification but likely compromised $D_{\alpha}$ extraction. Second, the method is computationally expensive as it required numerical Fourier inversion $P_{\text{1D}}(\xi,t) \rightarrow P_{\text{1D}}(x,t)$ and subsequent Hankel-Laplace transform $P_{\text{geo}}(r,t) \rightarrow P_{\text{geo}}(h,s)$. Third, the model was based on previously published jump drivers $\phi_1(u) \sim \exp(-|u/u_{\text{BD}}|)/|u|^{1+\alpha}$ which resulted in a propagator function $\psi_1(\xi)$ with cumbersome functional form.
\par
The framework presented here eliminates these shortcomings. First, a simple variable change $\xi_x \rightarrow \| \vec{\xi}\|$ in the propagator function rigorously extends a 1D Poissonian process to isotropic multidimensional transport. Second, the ability to operate fully within transformed domains boosts the computational efficiency to the extent that quasiballistic TDTR analysis (illustrated in Sec. \ref{sec:applications}.B) runs just as fast as conventional diffusive identification. Third, we replaced the bottom-up approach adopted by prior literature by a pragmatic one that describes transitions between asymptotic regimes directly within the propagator function in highly streamlined form.
\section{Application examples}\label{sec:applications}
\subsection{TTG analysis of GaAs}
TTG experiments use interference of two laser beams to subject the sample to a heating pulse that is spatially periodic (grating period $\lambda$) in one in-plane direction ($x$) \cite{TTG-fulldetails}. The induced power density $p$ is assumed to extend uniformly across the other in-plane direction ($y$) and decays exponentially in the cross-plane direction ($z$) with the optical penetration depth $d_{\text{opt}}$ of the pump laser:
\begin{eqnarray}
p(\vec{r},t) & \propto & \cos \left( 2\pi x/\lambda \right) \, \exp \left( - |z|/d_{\text{opt}}\right) \, \delta(t)\\
\leftrightarrow \, p(\vec{\xi},s) & \propto & \frac{\left[ \delta \left(\xi_x - \frac{2\pi}{\lambda} \right) + \delta \left(\xi_x + \frac{2\pi}{\lambda} \right) \right] \, \delta(\xi_y)}{1 + \xi_z^2 \, d_{\text{opt}}^2} \qquad \label{TTGsource}
\end{eqnarray}
Inverse transform of $(2/C_v) \, p(\vec{\xi},s) \, P(\vec{\xi},s)$ provides the surface temperature $\Delta T_0 \equiv \Delta T(x,y,z=0;t)$:
\begin{equation}
\Delta T_0 \propto \cos \left( \frac{2\pi x}{\lambda} \right) \, \int \limits_{0}^{\infty} \frac{P \left( \xi_x = \frac{2\pi}{\lambda}, \xi_y = 0, \xi_z;t \right) \, \mathrm{d}\xi_z}{1 + \xi_z^2 \, d_{\text{opt}}^2}
\end{equation}
The measured signal $\mathcal{M}$ consists of the time decay of the peak-to-valley contrast within $\Delta T_0$. Note that our reasoning deviates from the result given by Minnich \cite{minnich-TTG3D}, who incorporated an additional factor $(1+\xi_z^2 \, d'^2_{\text{opt}})^{-1}$ to account for cross-plane weighing of the thermal field by the probe beam. However, we argue that no such weighing takes place since the probe reflection is normally considered to originate at the actual sample surface \cite{TTG-fulldetails}. Modelling the thermal transport with a 3D isotropic Poissonian process $P(\zeta,t) = \exp[-\psi_3(\zeta) \, t]$ we obtain
\begin{equation}
\mathcal{M}(t;\lambda) \propto \int \limits_{0}^{\infty} \frac{\exp \left[ -\psi_3 \left( \zeta = \sqrt{(2\pi/ \lambda)^2 + \xi_z^2} \right) t \right] \, \mathrm{d}\xi_z}{1 + \xi_z^2 \, d_{\text{opt}}^2}\label{TTGsignal}
\end{equation}
The experiment is only sensitive to a limited spatial bandwidth of the thermal response: low spatial frequencies $\zeta < 2\pi/\lambda$ are not probed at all while high frequency contributions $\zeta > 1/d_{\text{opt}}$ are increasingly attenuated. This sets some limitations to the reconstruction of the initial portions of the phonon MFP spectrum \cite{minnich-TTG3D}. For our purposes, we should expect that reliable identification of the propagator function $\psi_3(\zeta)$ is achievable over the `critical window' $2\pi/\lambda_{\text{max}} \leq \zeta \leq 1/d_{\text{opt}}$.
\par
One can verify that with a Brownian propagator $\psi(\zeta) = D\, \zeta^2$, (\ref{TTGsignal}) correctly reproduces the analytic diffusive solution derived in Ref. \onlinecite{TTG-fulldetails}. For perfect surface heating ($d_{\text{opt}} \rightarrow 0$) the diffusive response obeys $\mathcal{M} \propto (Dt)^{-1/2} \, \exp ( -4 \pi^2 D t/\lambda^2)$; conventional analyses fit this simple form to the measured transients to extract an effective diffusivity $D_{\text{eff}}(\lambda)$ \cite{TTG-membranes, TTG-GaAs}.
\par
Here, we instead analyse raw TTG signal transients on bulk GaAs courtesy of Johnson and coworkers \cite{TTG-GaAs} with a log-tempered Cauchy propagator (\ref{logtemperedcauchypropagator}). The measurements were performed at a 535$\,$nm pump laser wavelength, and we accordingly set $d_{\text{opt}} = 170\,$nm based on the optical absorption curve for high-purity GaAs \cite{GaAs-optical}. The theoretical signals (\ref{TTGsignal}), easily evaluated numerically with a simple quadrature scheme, show good agreement with the experimental data for best-fitting values $D = 24\,$mm$^2$/s ($\kappa \simeq 41.5\,$W/m-K), $r_{\text{CF}} = 467\,$nm and $b=1.06$ (Fig. \ref{fig1_TTGsignals}).
\myfig[!htb]{width=0.45\textwidth}{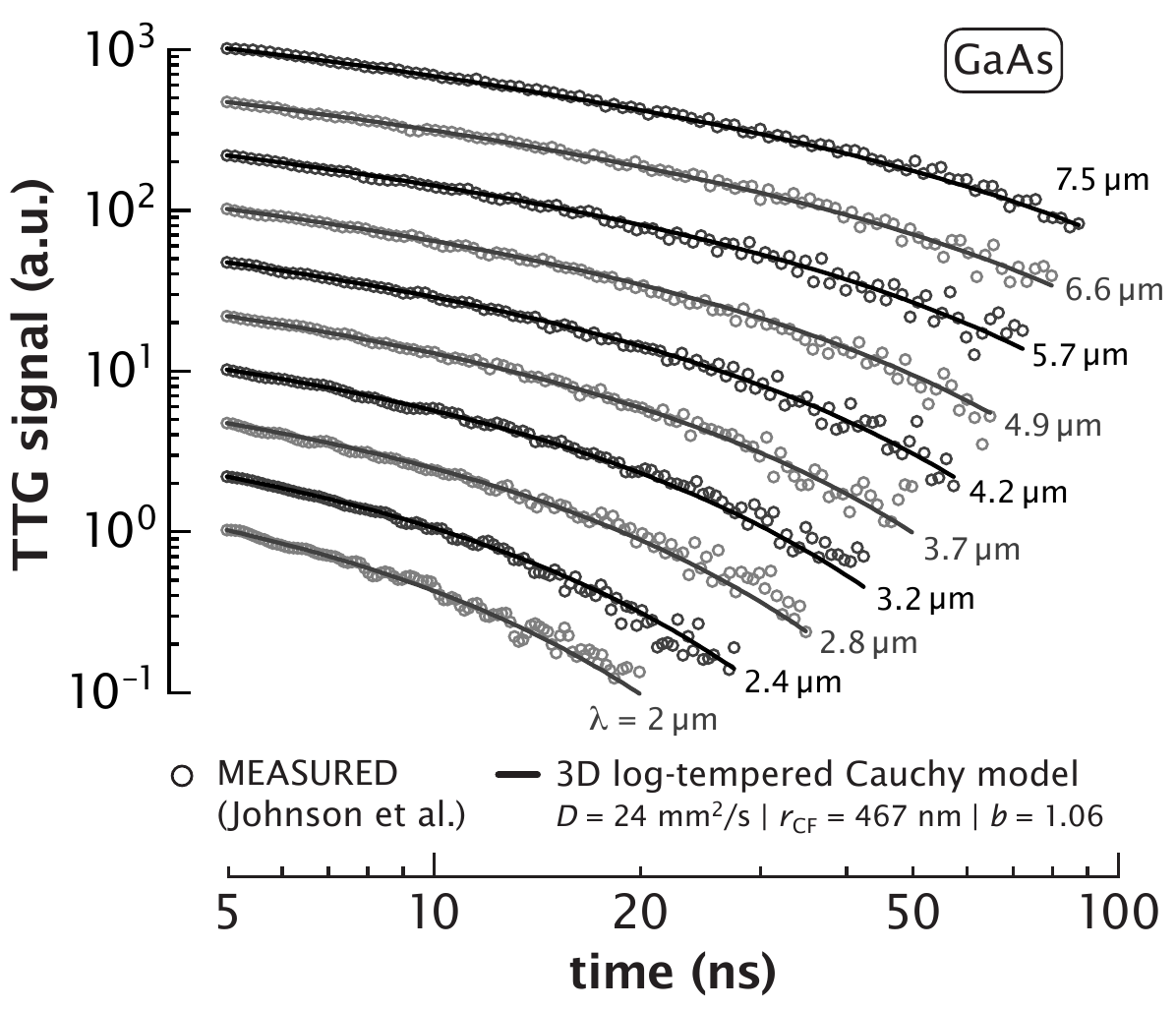}{Quasiballistic TTG analysis of bulk GaAs using a 3D isotropic Poissonian process with log-tempered Cauchy propagator $\psi_3(\zeta) = D \, \zeta^2 \, \ln [ 1 + (1 + r_{\text{CF}}^{b} \, \zeta^{b} )^{-1/b} ]/\ln(2)$. Experimental data courtesy of Johnson and coworkers. Curves were offset vertically for visual clarity.}
\par
Since $\lambda/r_{\text{CF}} > 4$ even at the smallest grating period, the measurements only probed the initial parts of the quasiballistic transition without having access to the pure Cauchy regime. As a consequence, the inferred Cauchy velocity $\bar{v} \simeq 74\,$m/s should not be taken at face value, and it indeed compares poorly with first-principles predictions ($\bar{v} \simeq 270\,$m/s).
\par
We have additionally found that a pure L\'evy propagator (with $\alpha = 1.563$ and $D_{\alpha} = 427\,$mm$^{\alpha}$/s) and tempered L\'evy propagator (with $D = 23.1\,$mm$^2$/s, $\alpha = 1.458$ and $r_{\text{LF}} = 936\,$nm) also provide good quality fits to the experimental data. We stress, however, one should not regard these observations as a proof of characteristic alloy behaviour in a single-crystal material. L\'evy dynamics in alloys form a genuine distinct transport regime with stable fractional exponent that persists across 2--3 orders of magnitude of spatial scale \cite{PRBlevy1,APLthinfilms}. A L\'evy fit to the GaAs TTG data, by contrast, merely constitutes a linear curve approximation in logarithmic coordinates: an initially parabolic function $\psi(\zeta \rightarrow 0) \sim \zeta^2$ that evolves to a linear asymptote $\psi(\zeta \rightarrow \infty) \sim \zeta$ can indeed be reasonably approximated over a sufficiently narrow $\zeta$ window by a fractional power law $\psi \sim \zeta^{\alpha}$ with intermediate exponent $1 < \alpha < 2$. Figure \ref{fig2_GaAs_propagators} illustrates that the three functional forms we have determined are conceptually quite different but, crucially, are indeed very similar inside the critical spatial frequency window.
\myfig[!htb]{width=0.45\textwidth}{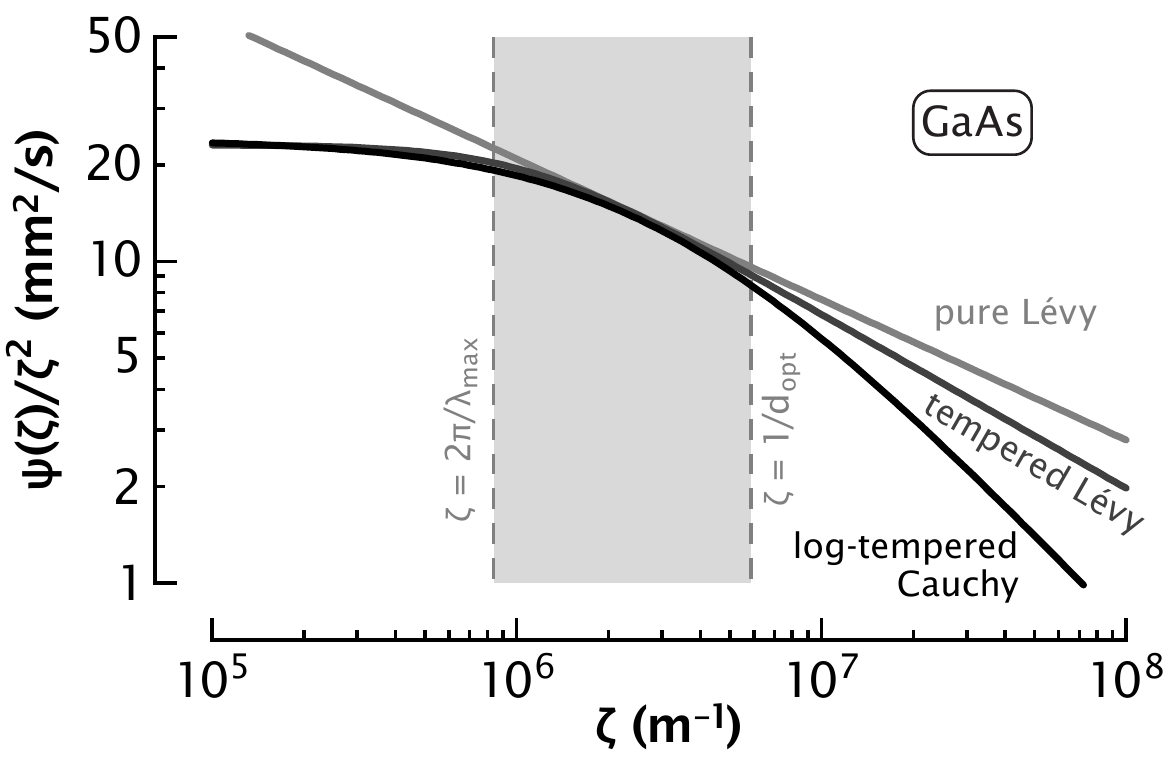}{Conceptually different propagators each providing a good fit to TTG experiments on GaAs. Reliable $\psi(\zeta)$ reconstruction is achievable over the critical window (shaded).}
\subsection{TDTR analysis of InGaAs}\label{sec:TDTR}
TDTR experiments subject the semiconductor under study, covered by a thin metal transducer, to a `pump' laser pulse train that is modulated at temporal frequency $f_{\text{mod}}$. Lock-in detection of the transducer surface reflectivity as monitored by a `probe' pulse train serves as basis for the thermal characterisation \cite{cahillmodel,cahill}. The presence of the transducer (and associated thermal resistivity $R_{\text{ms}}$ of the metal/semiconductor interface) as well as the complicated time signature of the heat source render TDTR analysis somewhat more involved than its TTG counterpart. First the single pulse temperature response of the transducer surface to a Gaussian pump beam is calculated in Hankel-Laplace domain with thermal quadrupoles \cite{quadrupoles}; the result is then weighed by the Gaussian probe beam and manipulated in temporal frequency domain to account for pulse repetition, modulation and lock-in detection \cite{cahillmodel}. Quasiballistic effects emerge when the thermal penetration length $\ell = \sqrt{D/\pi f_{\text{mod}}}$ becomes comparable with phonon MFPs \cite{cahill,PRBcomment}. Even so, conventional analyses still assume purely diffusive transport in the semiconductor and then extract a frequency-dependent effective thermal conductivity $\kappa_{\text{eff}}(f_{\text{mod}})$ \cite{cahill,PRBlevy2}.
\par
Replacing the Green's function of the semiconductor surface $G_0(h,s) \equiv (2/C_v)\, P(h,z=0,s)$ [the factor 2 accounts for the semi-infinite geometry] by a suitable quasiballistic expression enables a more refined characterisation. For isotropic Poissonian flight dynamics we have
\begin{equation}
G_{0}(h,s) = \frac{2}{\pi \, C_v} \int \limits_{0}^{\infty} \frac{\mathrm{d}\xi_z}{s + \psi_3(\zeta = \sqrt{h^2 + \xi_z^2})}
\end{equation}
A numerical scheme is easily devised by observing that for a piecewise linear Taylor expansion of the propagator function we have $\int \mathrm{d}\xi_z/(s+A+B\,\xi_z) = B^{-1}\, \ln(s+A+B\,\xi_z) \simeq (\partial \psi_3/\partial \xi_z)^{-1} \, \ln(s + \psi_3)$, hence
\begin{equation}
G_0(h,s) \simeq \frac{2}{\pi \, C_v} \sum \limits_{n=1}^{N-1} \left [ \frac{\ln \left( \frac{s+\psi^{(n+1)}}{s+\psi^{(n)}}\right)}{\psi^{(n+1)} - \psi^{(n)}} \right] \cdot \Delta \xi_z^{(n)} \label{Gsurfnumerical}
\end{equation}
where $\psi^{(i)} \equiv \psi_3(h,\xi_z^{(i)})$ and $\Delta \xi_z^{(n)} \equiv \xi_z^{(n+1)} - \xi_z^{(n)}$. We note that if $\psi^{(n)} = \psi^{(n+1)}$ within machine precision, which routinely occurs for $\xi_z \ll h$, the expression between square brackets in (\ref{Gsurfnumerical}) must be replaced by $[s + \psi^{(n)}]^{-1}$. A logarithmic $\xi_z$ grid with 500 points usually suffices for accurate computation; the numerical result for a tempered L\'evy propagator with $\alpha = 1.999$ and $r_{\text{LF}} = 1\,$nm matched the magnitude and phase of the exact diffusive solution $G_0 = \kappa^{-1} \, (s/D + h^2)^{-1/2}$ within 0.09\% and 0.02 degrees respectively.
\par
We use our framework with 3D tempered L\'evy propagator (\ref{temperedlevypropagator}) to analyse raw TDTR signal transients recorded on an In$_{0.53}$Ga$_{0.47}$As sample ($C_v = 1.546\,$MJ/m$^3$-K) with 64$\,$nm Al transducer (Fig. \ref{fig3_TDTRtemperedlevy}).
\myfig[!htb]{width=0.45\textwidth}{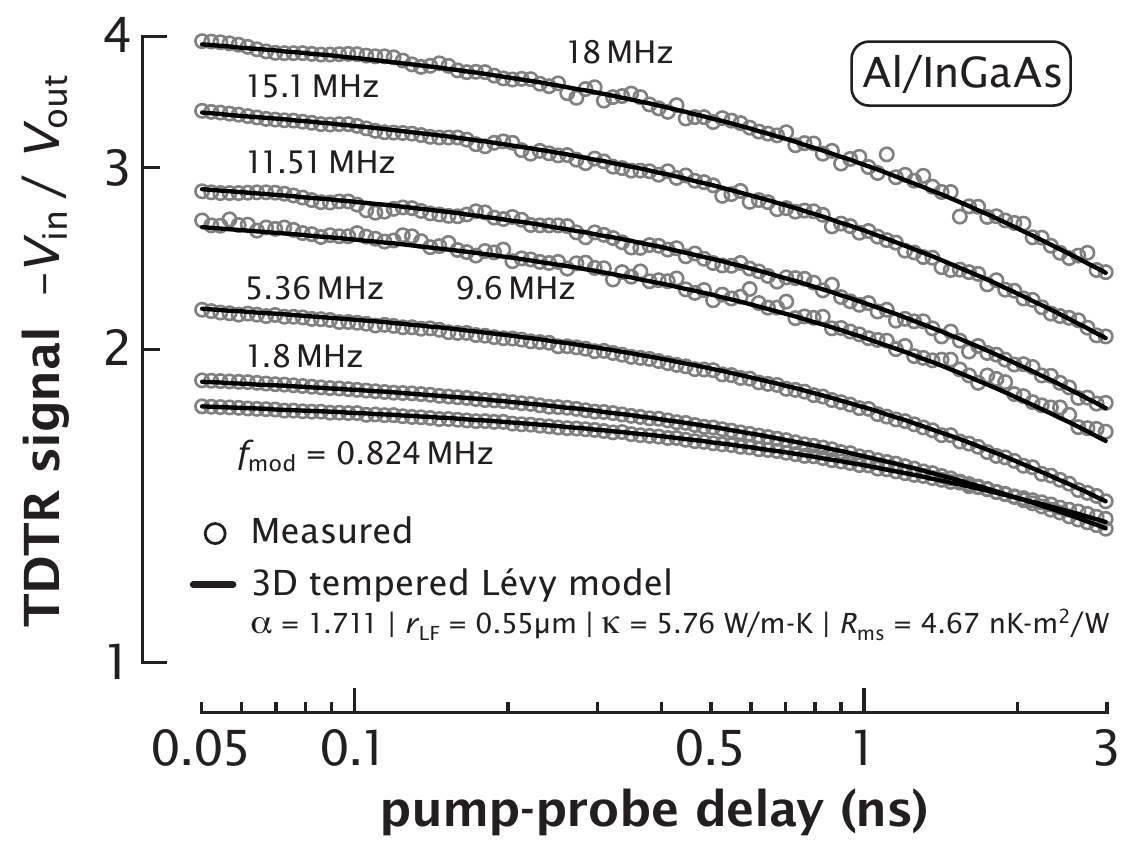}{Quasiballistic TDTR analysis of InGaAs sample with Al transducer using a 3D isotropic Poissonian process with tempered L\'evy propagator $\psi(\zeta) = D \, \zeta^2/(1+r_{\text{LF}}^2\, \zeta^2)^{1-\alpha/2}$.}
\par
From collective least-square optimisation on all available curves we find best-fitting values for a total of 4 model parameters: L\'evy exponent $\alpha = 1.711$ (+2\%); L\'evy-Fourier transition length $r_{\text{LF}} = 0.55\,\mu$m; nominal Fourier conductivity $\kappa = 5.76\,$W/m-K ($-$3\%); and Al/InGaAs interface resistivity $R_{\text{ms}} = 4.67\,$nK-m$^2$/W (+13\%). Bracketed values indicate relative changes from results previously obtained by our original `truncated L\'evy' model \cite{PRBlevy2}. The transition length $r_{\text{LF}}$ cannot be directly compared to our prior truncated L\'evy result $u_{\text{BD}} = 3.36\,\mu$m since the latter metric has a different quantitative meaning and was extracted using a geometrically extended 1D process. What matters is that the metrics should provide roughly consistent representations of the same actual semiconductor response. Equating the prefactors in the L\'evy source response $G_{0}(r=0,t) \sim t^{-3/\alpha}$ with $\alpha = 1.711$ for both models gives a theoretical mapping ratio $r_{\text{LF}}/u_{\text{BD}} \approx 0.18$, indeed quite close to the value $0.55/3.36 \approx 0.16$ observed in practice.
\par
Having $\ell \leq r_{\text{LF}}$ for $f_{\text{mod}} \geq 3.92\,$MHz suggests that the experiment probes fairly deeply into the alloy's L\'evy regime, as predicted from first principles \cite{PRBlevy1}. We can verify this directly by analysing the signals with a pure L\'evy propagator $\psi(\zeta) = D_{\alpha} \, \zeta^{\alpha}$. Excellent performance is observed for the upper 4 modulation frequencies with best fitting values $\alpha = 1.703$, $D_{\alpha} = 35.01\,$mm$^{1.703}$/s and $R_{\text{ms}} = 4.17\,$nK-m$^2$/W (Fig. \ref{fig4_TDTRpurelevy}).
\myfig[!htb]{width=0.45\textwidth}{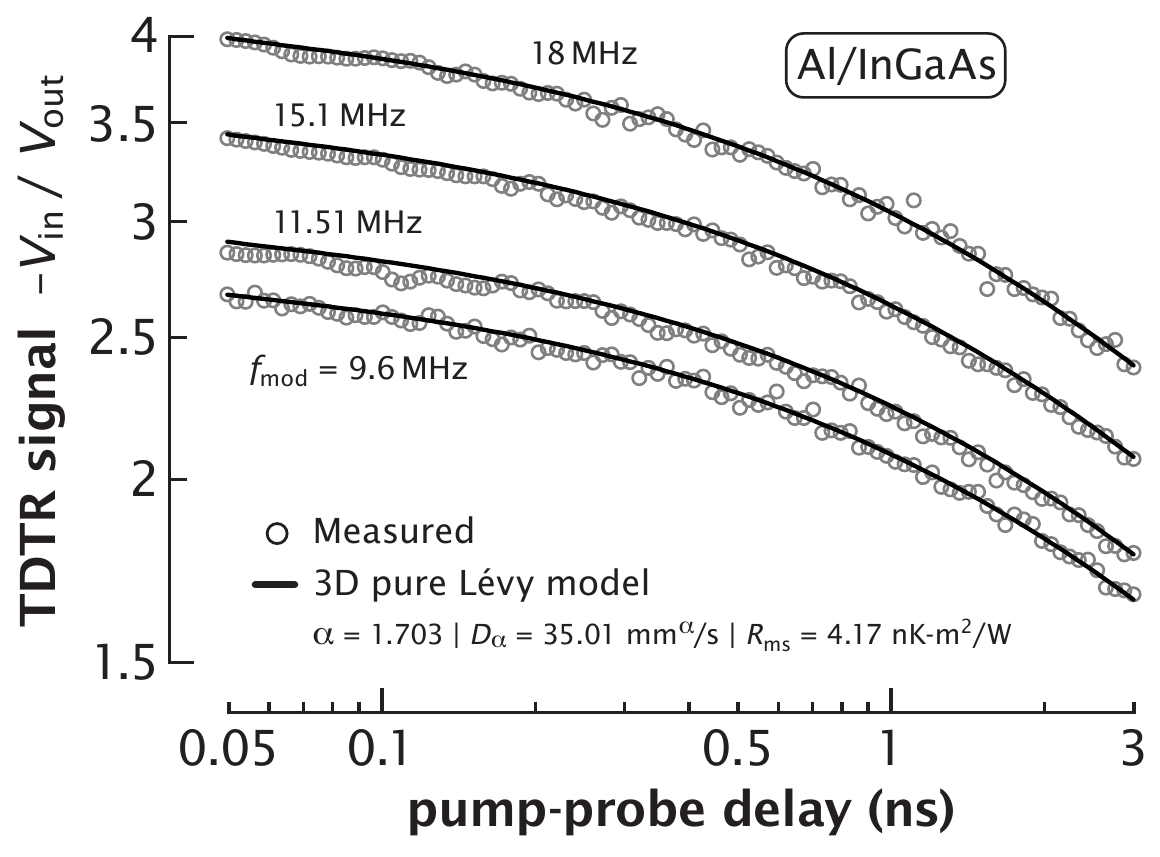}{Quasiballistic TDTR analysis at high modulation frequencies of InGaAs sample using a 3D isotropic Poissonian process with pure L\'evy propagator $\psi(\zeta) = D_{\alpha} \, \zeta^{\alpha}$.} 
\par
If access to the pure L\'evy regime is available, one may consider a two-tier fitting strategy. First, $\alpha$ and $R_{\text{ms}}$ can be determined from high modulation frequency data (where sensitivities are highest) through the pure L\'evy model. Then, the extracted values can be fixed in subsequent tempered L\'evy analysis across all available frequencies to identify the nominal conductivity and L\'evy-Fourier transition length. For our InGaAs example this yielded $\kappa = 5.77\,$W/m-K and $r_{\text{LF}} = 0.53\,\mu$m, in close agreement with the results obtained from single-tier tempered L\'evy analysis. The two-tier approach is also internally consistent: the fractional diffusivity inferred by the second stage, $D_{\alpha} = (\kappa/C_v)/r_{\text{LF}}^{2-\alpha} \simeq 35.06\,$mm$^{1.703}$/s, deviates less than 0.2\% from the value found independently as free parameter by the first stage.
\section{Conclusions}\label{sec:summary}
In summary, we presented a multidimensional analysis framework for nondiffusive thermal transport based on stochastic theory of isotropic Poissonian flight processes. The approach offers comprehensive characterisation of the quasiballistic heat dynamics beyond conventional `effective Fourier' interpretations with minimal computational overhead.
\section*{Acknowledgements}
The author acknowledges funding from the \textsc{alma} Horizon 2020 project (European Union Grant No. 645776) and thanks Jeremy Johnson (BYU) and Alexei Maznev (MIT) for sharing their TTG experiment data. The TDTR signals and suggestion of relativistic alpha-stable processes are courtesy of respectively UC Santa Cruz (Gilles Pernot, Ali Shakouri) and Samy Tindel (Purdue).
\appendix
\section{Isotropic Fourier transforms}\label{app:fouriertransforms}
The Fourier image $F$ of a $d$-dimensional radially symmetric function $f$ is generically given by \cite{fouriertransforms}:
\begin{equation}
F(\|\vec{\xi}\|) = (2\pi)^{d/2} \, \int \limits_{0}^{\infty} J_{\nu} (\|\vec{\xi}\| \, \|\vec{r}\|) \frac{\|\vec{r}\|^{\nu + 1}}{\|\vec{\xi}\|^{\nu}} f(\|\vec{r}\|) \, \mathrm{d}\|\vec{r}\| \label{fouriergeneric}
\end{equation}
Here $J_{\nu}$ denotes the Bessel function of the first kind of order $\nu \equiv (d-2)/2$. The inverse transform obeys the same formula with an additional prefactor $(2\pi)^{-d}$. Using dedicated notations $\xi \equiv |\xi_x|$, $h \equiv (\xi_x^2 + \xi_y^2)^{1/2}$, and $\zeta \equiv (\xi_x^2 + \xi_y^2 + \xi_z^2)^{1/2}$ for convencience, we have
\begin{eqnarray}
\text{1D : } F(\xi) & = & 2 \, \int \limits_{0}^{\infty} \cos(\xi x) \, f(x) \, \mathrm{d}x \label{FFT1D} \\
\leftrightarrow \quad f(x) & = & \frac{1}{\pi} \, \int \limits_{0}^{\infty} \cos(\xi x) \, F(\xi) \, \mathrm{d}\xi \label{IFT1D}
\end{eqnarray}
which is indeed the standard Fourier transform specialised to an even function;
\begin{eqnarray}
\text{2D : } F(h) & = & 2\pi \, \int \limits_{0}^{\infty} J_0(h\rho) \, f(\rho) \, \rho \, \mathrm{d}\rho \label{FFT2D} \\
\leftrightarrow \quad f(\rho) & = & \frac{1}{2\pi} \, \int \limits_{0}^{\infty} J_0(h\rho) \, F(h) \, h \, \mathrm{d}h \label{IFT2D}
\end{eqnarray}
which is the familiar Hankel transform; and
\begin{eqnarray}
\text{3D : } F(\zeta) & = & 4\pi \, \int \limits_{0}^{\infty} j_0(\zeta r) \, f(r) \, r^2 \, \mathrm{d}r \label{FFT3D} \\
\leftrightarrow \quad f(r) & = & \frac{1}{2\pi^2} \, \int \limits_{0}^{\infty} j_0(\zeta r) \, F(\zeta) \, \zeta^2 \, \mathrm{d}\zeta \label{IFT3D}
\end{eqnarray}
where we expressed the results in terms of the zeroth-order spherical Bessel function $j_0(\zeta r) \equiv \mathrm{sinc}(\zeta r)$.
\section{Rigorous connection between isotropic multidimensional processes and 1D counterparts}\label{app:mappingproof}
Consider a $d$-dimensional isotropic stochastic process $\mathcal{S}$ with single pulse response $P_d(\vec{r},t)$. Let $\hat{\mathcal{S}}_1$ be the projection $\mathbb{R}^d \rightarrow \mathbb{R}$ of $\mathcal{S}$ onto a line through the origin, which without loss of generality can be made the cartesian $x$-axis. The single pulse response of $\hat{\mathcal{S}}_1$, which we will denote by $\hat{P}_{d \rightarrow 1}(x,t)$, constitutes the marginal density of $\mathcal{S}$ with respect to $x$ and thus follows from integrating $P_d(\vec{r},t)$ over the other $d-1$ space coordinates. Here we explicitly demonstrate for $d=2$ and $d=3$ that the Fourier images $\hat{P}_{d \rightarrow 1}(\xi_x,t)$ and $P_d(\vec{\xi},t)$ are formally identical except for straightforward variable exchange $\xi_x \leftrightarrow \|\vec{\xi}\|$. In turn, propagator functions obey the same symmetry, because all spatial dependencies of Poissonian processes are contained therein.
\subsection*{Mapping between 2D and 1D}
Here we have
\begin{equation}
\hat{P}_{2 \rightarrow 1}(x,t) = 2 \int \limits_{0}^{\infty} P_2(\rho = \sqrt{x^2+y^2},t) \, \mathrm{d}y
\end{equation}
where we used the evenness of $P_2$ with respect to $y$. Taking the 1D Fourier transform of both sides yields
\begin{equation}
\hat{P}_{2 \rightarrow 1}(\xi,t) = 4 \int \limits_{0}^{\infty} \mathrm{d}x \, \cos(\xi x) \, \int \limits_{0}^{\infty} \mathrm{d}y \, P_2(\rho = \sqrt{x^2+y^2},t)
\end{equation}
Reverting to cylindrical coordinates produces
\begin{eqnarray}
\hat{P}_{2 \rightarrow 1}(\xi,t) & = & 4 \int \limits_{0}^{\pi/2} \mathrm{d} \varphi \, \cos(\xi \rho \cos \varphi) \, \int \limits_{0}^{\infty} \rho \, \mathrm{d}\rho \, P_2(\rho,t) \nonumber \\
& = & 2\pi \int \limits_{0}^{\infty} \rho \, J_0(\xi \rho) \, P_2(\rho,t) \, \mathrm{d}\rho
\end{eqnarray}
The latter is, by definition, the Hankel transform of $P_2$ evaluated at $h=\xi$, so we arrive at
\begin{equation}
\hat{P}_{2 \rightarrow 1}(\xi,t) \equiv P_2(\sqrt{\xi_x^2 + \xi_y^2}=\xi,t)
\end{equation}
\subsection*{Mapping between 3D and 1D}
In analogy to the 2D case just discussed, we have
\begin{equation}
\hat{P}_{3 \rightarrow 1}(x,t) = 4 \int \limits_{0}^{\infty} \mathrm{d}y \, \int \limits_{0}^{\infty} \mathrm{d}z \, P_3(r = \sqrt{x^2+y^2+z^2},t)
\end{equation}
Taking the 1D Fourier transform of both sides and reverting to spherical coordinates results in
\begin{multline}
\hat{P}_{3 \rightarrow 1}(\xi,t) = \\
8 \int \limits_{0}^{\pi/2} \mathrm{d} \varphi \, \cos(\xi r \cos \varphi \, \sin \theta) \, \int \limits_{0}^{\pi/2} \sin \theta \, \mathrm{d}\theta \, \int \limits_{0}^{\infty} r^2 \, \mathrm{d}r \, P_3(r,t)
\end{multline}
The integral over $\varphi$ produces $(\pi/2) \, J_0(\xi r \sin \theta)$. Subsequent $\theta$ integration gives $(\pi/2) \, j_0(\xi r)$, so we arrive at
\begin{equation}
\hat{P}_{3 \rightarrow 1}(\xi,t) = 4 \pi \int \limits_{0}^{\infty} r^2 \, j_0(\xi r) \, P_3(r,t) \, \mathrm{d}r
\end{equation}
Here we recognise the spherical Bessel transform of $P_3$ evaluated at $\zeta=\xi$, hence
\begin{equation}
\hat{P}_{3 \rightarrow 1}(\xi,t) \equiv P_3(\sqrt{\xi_x^2 + \xi_y^2 + \xi_z^2}=\xi,t)
\end{equation}
\subsection*{Corollary: mapping between 3D and 2D}
Combining the two mapping results just proven immediately shows that
\begin{equation}
\hat{P}_{3 \rightarrow 2}(h,t) \equiv P_3(\zeta=h,t)
\end{equation}
One can also verify this explicitly by taking the marginal density of $P_3(r = [\rho^2 + z^2]^{1/2},t)$ with respect to $\rho$ and then calculating its Hankel transform.
%
\end{document}